\documentclass[11pt]{article}
\usepackage{graphicx}
\usepackage{amssymb}
\usepackage[margin=1.25in]{geometry}
\usepackage[usenames,dvipsnames]{color}
\usepackage{url}
\usepackage[colorlinks = true,
            linkcolor = blue,
            urlcolor  = blue,
            citecolor = blue,
            anchorcolor = blue]{hyperref}

\newcommand\pubnumber{   }
\newcommand\pubdate{\today}

\def\Title#1{\begin{center} {\LARGE #1 } \end{center}}
\def\Author#1{\begin{center}{ \sc #1} \end{center}}
\def\Address#1{\begin{center}{ \it #1} \end{center}}

\newcommand\pubblock{\rightline{\begin{tabular}{l} \pubnumber\\
         \pubdate \end{tabular}}}
\newenvironment{Abstract}{\begin{quotation} \begin{center}
                       ABSTRACT
     \end{center}\bigskip  }{\end{quotation}}

\def\beq{\begin{equation}}
\def\eeq#1{\label{#1}\end{equation}}
\def\eeqn{\end{equation}}

\newenvironment{Eqnarray}%
   {\arraycolsep 0.14em\begin{eqnarray}}{\end{eqnarray}}
\def\beqa{\begin{Eqnarray}}
\def\eeqa#1{\label{#1}\end{Eqnarray}}
\def\eeqan{\end{Eqnarray}}

\let\bar=\overbar

\def\lsim{\mathrel{\raise.3ex\hbox{$<$\kern-.75em\lower1ex\hbox{$\sim$}}}}
\def\gsim{\mathrel{\raise.3ex\hbox{$>$\kern-.75em\lower1ex\hbox{$\sim$}}}}

\def\del{\partial}
\def\Dslash{\not{\hbox{\kern-4pt $D$}}}
\def\dslash{\not{\hbox{\kern-2pt $\del$}}}
\def\pslash{\not{\hbox{\kern-2pt $p$}}}
\def\ETmiss{\not{\hbox{\kern-4pt $E$}}_T}

\def\Dlr{\mathrel{\raise1.5ex\hbox{$\leftrightarrow$\kern-1em\lower1.5ex\hbox{$D$}}}}

\def\MSB{{\bar{M \kern -2pt S}}}
\def\msb{{\bar{\scriptsize M \kern -1pt S}}}

\def\drb{{\bar{\scriptsize D \kern -1pt R}}}

\newcommand\snowmass{\begin{center}\rule[-0.2in]{\hsize}{0.01in}\\\rule{\hsize}{0.01in}\\
\vskip 0.1in Submitted to the  Proceedings of the US Community Study\\ 
on the Future of Particle Physics (Snowmass 2021)\\ 
\rule{\hsize}{0.01in}\\\rule[+0.2in]{\hsize}{0.01in} \end{center}}

\begin{document}

\pubblock

\Title{Top-quark mass extraction from $t\bar{t}j +X$ events at the LHC: theory predictions}


\Author{Simone Alioli$^{1}$, Juan~Fuster$^{2}$,
Maria~Vittoria~Garzelli$^{3}$, Alessandro~Gavardi$^{1}$,
Adrian~Irles$^{2}$, 
Davide~Melini$^{4}$,
Sven-Olaf~Moch$^{3}$,
Peter~Uwer$^{5}$,
Katharina~Vo\ss$^{6}$}


\Address{
$^1$ Dipartimento di Fisica “G. Occhialini”, Universit\`a degli Studi di Milano-Bicocca, and INFN, Sezione di Milano Bicocca, Piazza della Scienza 3, I~--~20126 Milano, Italy\\
$^2$ IFIC, Universitat de Val\`encia and CSIC,
  Catedr\'atico Jose Beltr\'an 2, E~--~46980 Paterna, Spain\\
$^3$ II. Institut f\"ur Theoretische Physik, Universit\"at
  Hamburg, Luruper Chaussee 149, D~--~22761 Hamburg, Germany\\
$^4$ Department of Physics, Technion, 
 Israel Institute of Technology, Haifa, Israel\\
$^5$ Institut f\"ur Physik, Humboldt-Universit\"at zu
  Berlin, Newtonstra{\ss}e 15, D~--~12489 Berlin, Germany\\
$^6$ Center for Particle Physics, Department f\"ur Physik, Universit\"at Siegen, Emmy Noether Campus, Walter Flex Str. 3, D~--~57068 Siegen, Germany
}

\medskip

\begin{Abstract}
\noindent 
Past work has proven the possibility of extracting the top-quark mass, one of the fundamental parameters of the Standard Model, from the comparison of theory predictions and experimental measurements of differential cross-sections for $t\bar{t} j~+~X$ hadroproduction. Various experimental analyses in this respect have  already been performed, and new ones are in preparation on the basis of the latest data from $pp$ collisions collected at the Large Hadron Collider. We have produced and made public a comprehensive set of theoretical predictions for the relevant differential distributions, ready to be used for presently ongoing and forthcoming experimental analyses. We investigate the role of different theoretical inputs, in particular the factorization and renormalization scales, PDFs and top-quark mass renormalization schemes, and we quantify the uncertainties related to different choices for these inputs, providing recommendations. 
\end{Abstract}

\snowmass

\def\thefootnote{\fnsymbol{footnote}}
\setcounter{footnote}{0}

\section{Introduction}
\label{sec:intro}

The masses of the heavy quarks are fundamental input parameters of the Standard Model (SM). The top-quark mass represents a particularly interesting case, due to its largeness, closeness to the electroweak scale, key role played in the SM precision tests and relation to the stability of the SM vacuum~\cite{Corcella:2019tgt, Hoang:2020iah}. For all these reasons, many analyses have focused on the extraction of the top-quark mass, using different direct and indirect methods. Among the latter, we consider here in particular the possibility of determining the top-quark mass by comparing experimental data for specific differential cross-sections with the corresponding theory predictions. One of the advantages of this method is the possibility to extract the top-quark mass in well defined mass renormalization schemes, used to perform the theoretical calculations of the cross-sections. A distribution that turned out to be particularly sensitive to the top-quark mass value, and thus is well suited for its determination, is the so-called $\mathcal{R}$ distribution in $t\bar{t}j + X$ hadroproduction events~\cite{Alioli:2013mxa, Fuster:2017rev}. $\mathcal{R}$ is built from the $\rho_s$ distribution, which, in turn, is inversely proportional to the invariant mass of the $t\bar{t}{j}$ system $s_{t\bar{t}j}$, i.e.
\begin{equation}
  {\cal R}(m_t^R,\rho_s)= 
  \frac{1}{{\ensuremath{\sigma_{t\bar t + \textnormal{\scriptsize 1-jet}}}}} 
  \frac{d{\ensuremath{\sigma_{t\bar t + \textnormal{\scriptsize 1-jet}}}}}{d\rho_s}(m_t^R, \rho_s), \,\,\, \mathrm{with}\,\,\,
  \rho_s \,=\, \frac{2 m_0} {\sqrt{{\ensuremath{s_{t\bar t j}}}}} \, .
\end{equation}
In the previous equation $m_t^R$ denotes the top-quark mass value in the $R$ renormalization scheme, whereas $m_0$ is a constant parameter of the order of the top-quark mass itself (we fix $m_0$ to the value 170 GeV throughout this work). It has been shown that the shape of the $\mathcal{R}$ distribution is extremely sensitive to $m_t^R$ and that this sensitivity increases when using samples of $t\bar{t}j +X $ events instead of samples of $t\bar{t} + X$ events. Events for the latter process, in fact, could also be used for building a $\rho_s$ and $\mathcal{R}$ distribution, with definition similar to Eq.~(1) but replacing $s_{t\bar{t}j}$ with $s_{t\bar{t}}$ and $\sigma_{t\bar{t} + \textnormal{\scriptsize 1-jet}}$ with $\sigma_{t\bar{t}}$.  
In the following we focus on the $\rho_s$ and $\mathcal{R}$ distributions for $t\bar{t}j + X$ production in $pp$ collisions at the Large Hadron Collider (LHC). These distributions were already used in some experimental analyses for top-quark mass extraction~\cite{ATLAS:2015pfy, CMS:2016khu, ATLAS:2019guf} with data from collisions at $\sqrt{S} = 7$ and 8~TeV and are still used in further ongoing analyses with data at $\sqrt{S} = 13$~TeV. 
We present theoretical predictions for them and the associated theoretical uncertainties. In this study, summarizing the more extended one we presented in Ref.~\cite{Alioli:2022lqo}, we include QCD radiative corrections at next-to-leading order (NLO)  and we focus on the case of stable top quarks. In fact, the experimental collaborations have developed sophisticated methods to reconstruct the top quarks from their decay products and they indeed apply these techniques in their $t\bar{t}j + X$ analyses devoted to $m_t^R$ extraction~\cite{ATLAS:2019guf, CMS:2019esx}.

\section{Theoretical framework}
\label{sec:theo}

In order to produce predictions for $t\bar{t}j + X$ production in $pp$ collisions, we consider two different and independent frameworks including NLO QCD radiative corrections: the most updated version of the implementation of Ref.~\cite{Dittmaier:2007wz, Dittmaier:2008uj}, making use of virtual amplitudes computed analytically with tensor reduction techniques, and a fully numerical approach, implemented in the \texttt{POWHEG-BOX-v2} framework~\cite{Alioli:2010xd}, using as input virtual amplitudes generated fully numerically by the {\texttt{OpenLoops-v2}} code~\cite{Buccioni:2019sur}. We checked that the two implementations produce NLO QCD predictions in perfect agreement among each other. Various other frameworks exist nowadays which allow to perform computations of the cross-sections of this process with the same accuracy. Implementations also exist capable of accounting for off-shellness and spin-correlation effects in the description of the top-quark production and decay (see e.g. Ref.~\cite{Bevilacqua:2015qha, Bevilacqua:2016jfk}) However, considering that the experimental analyses reconstruct top quarks from their decay and further emission products, and then extract top-quark masses from events with reconstructed top-quarks, we did not need to adopt these implementations for producing the predictions in this work. In fact the theoretical results from these techniques have already been considered and used to test the robustness and further refine the top-quark reconstruction techniques. This has already happened in the framework of some of the analyses (see e.g. the studies we presented in section 4.4 of Ref.~\cite{Alioli:2022lqo} and references therein). We expect that this way of proceeding will be extended even to future  analyses. The results we have obtained so far have shown that off-shell and spin-correlation effects do not produce modifications of the $\mathcal{R}$ distribution substantial enough to induce relevant shifts in the value of the extracted top-quark mass, with respect to the case where these effects are neglected. On the other hand, shower emissions turn out to be more relevant and they have to be taken fully into account in the reconstruction of the top quarks. The NLO QCD implementations with stable top quarks have 
been matched~\cite{Kardos:2011qa, Alioli:2011as, Alwall:2014hca, Czakon:2015cla} with the parton shower implementations available in different Shower Monte Carlo codes, by using different matching methods. Leading NLO EW corrections have also been incorporated and merging techniques, allowing to combine $t\bar{t}j$ event samples with samples with an higher multiplicity of jets have also been applied in the most advanced approaches~\cite{Gutschow:2018tuk}, combining both matching and merging. A comprehensive and up-to-date list of the available implementations and their features is reported in Ref.~\cite{Alioli:2022lqo}. On the other hand, 2-loop virtual corrections have not been calculated/computed yet, and, therefore, next-to-NLO (NNLO) theoretical predictions for total and differential cross-sections for the considered process are not yet available.

\subsection{Input of the theoretical computations and estimate of the related uncertainties}
\label{subsec:input}

Consistently with the accuracy of our fixed-order computation, we use as input state-of-the-art NLO PDFs from various collaborations, accompanied by their $\alpha_S(M_Z)$ value and $\alpha_S(M_Z)$ 2-loop evolution, as provided by the {\texttt{LHAPDF}} interface~\cite{Buckley:2014ana}. Alternatively, we also fix $\alpha_s(M_z)$ to a standard value (0.118) and evolve $\alpha_S$ at 2-loop via an in-house code. The number of active flavours is fixed to five at all scales relevant for the calculation.
By default, we present predictions for top-quark mass renormalized in the on-shell scheme. As alternatives, we also provide predictions for top-quark mass renormalized in the ${{\rm \overline{MS}}}$ and MSR schemes~\cite{Hoang:2008yj,Hoang:2017suc}. In case of MSR, we work in the so-called MSRp variant. The reason underlying the use of these different schemes is the fact that they lead to short-distance masses, i.e. masses free from the renormalon uncertainty, which is instead due to long-distance effects, and affects every on-shell quark mass measurement. This uncertainty, on the size of which the scientific community does not agree yet, but that amounts to at least $\mathcal{O}$(100 MeV)  according to the less conservative estimates, on the long term might be large enough to constitute a bottleneck for future top-quark mass measurements, which will be characterized by decreasing statistical and systematic uncertainties. 
QCD does not provide a univocal recipe to fix the renormalization and factorization scales $\mu_R$ and $\mu_F$, which are also input of our fixed-order calculation in collinear factorization. The central renormalization and factorization scales are assumed to be equal, i.e. $\mu_{R,0} = \mu_{F,0} = \mu_0$. We investigate the effect of various possible choices for $\mu_0$.   
Besides the static choice $\mu_0 = m_t^R$, according to which the scale is assumed equal to the value of the top-quark mass in the specific heavy-quark mass renormalization scheme considered for the calculation of the cross-section, when working in the on-shell scheme we also consider three additional choices, i.e.
$\mu_0 = H_T^B/2$, $\mu_0 = H_T^B/4$ and $\mu_0 = m_{t\bar{t}j}^B$, where 
$H_T$ denotes the sum of the transverse masses of the top quark, antitop quark and extra parton, $m_{t\bar{t}j}$ is the invariant mass of the system, and  
the superscript $B$ indicates that the corresponding quantities are computed using as input the underlying Born kinematics, i.e. the kinematics of the event before first radiation emission, as easily accessible within the \texttt{POWHEG-BOX-v2} implementation. 

On top of the aforementioned choices for PDFs, $\alpha_s(M_Z)$, $m_t^R$, $\mu_{R,0}$ and $\mu_{F,0}$, we also evaluate related uncertainties, using standard procedures. In particular, PDF uncertainties are evaluated by following the prescriptions specific to each set, according to the details specified by the corresponding PDF collaboration. $\alpha_S(M_Z)$ uncertainties are evaluated for the case of the ABMP16 NLO PDFs~\cite{Alekhin:2018pai}, by considering the existence of different PDF sets, each one corresponding to a different $\alpha_S(M_Z)$ value. This allows to fully account for the correlations between PDFs and $\alpha_S(M_Z)$, neglected in various other PDF fits.  
The top-quark mass value is varied on a wide enough range of values, 
in steps of 1~GeV, for all the considered renormalization schemes.
In the most general case, the scale uncertainty is evaluated by following the so-called seven-point prescription, i.e. by varying $\mu_R$ and $\mu_F$ by a factor [1/2,2] around their central values, building the envelope of the seven combinations \{(1,1), (1,2), (2,1), (2,2), (1,0.5), (0.5,1), (0.5, 0.5)\}($\mu_{R,0}$, $\mu_{F,0}$). While considering all these combinations is important in order to estimate the uncertainty affecting the $\rho_S$ distributions, it turns out that, for the $\mathcal{R}$ distribution, scale uncertainty is dominated by the simultaneous variation of $\mu_{R,0} = \mu_{F,0}$, i.e. by the combinations 
\{(1,1), (2,2),  (0.5, 0.5)\}($\mu_{R,0}$, $\mu_{F,0}$).

\subsection{Analysis cuts}
\label{subsec:cuts}

The considered process is already divergent at leading order (LO), meaning that phase-space cuts on the extra-parton are necessary to obtain a finite cross-section.
At NLO, events with one or two extra-partons can appear. A jet algorithm is applied, which leads to events with one or two jets, depending on whether the two partons are recombined together or not when applying the jet algorithm. The energy-recombination scheme is used for combining the momenta of different partons. The top quarks, being massive, are not part of any jet.
We apply analysis cuts similar to those applied in contemporary experimental analyses of $t\bar{t}j + X$ production aimed at extracting the top-quark mass, performed by both the ATLAS and CMS collaborations. An example of a typical setup, that we are going to consider in the following of this work, is selecting events with two top quarks (without cuts) and at least one extra jet, with transverse momentum $p_{T,j} > 30$ GeV and pseudorapidity $|\eta_j| < 2.4$, reconstructed with the anti-$k_T$ jet clustering algorithm~\cite{Cacciari:2008gp}  with $R=0.4$. 
We also produce predictions using tighter cuts, i.e. considering larger $p_{T,j}$ values (i.e. $p_{T,j} >$ 50, 70, 100 GeV), and for slightly different $|\eta_j|$ cuts (i.e. $|\eta_j| < 2.5$).

\section{Predictions}
\label{sec:pheno}

\begin{figure}[htb]
\begin{center}
\includegraphics[width=0.9\textwidth]{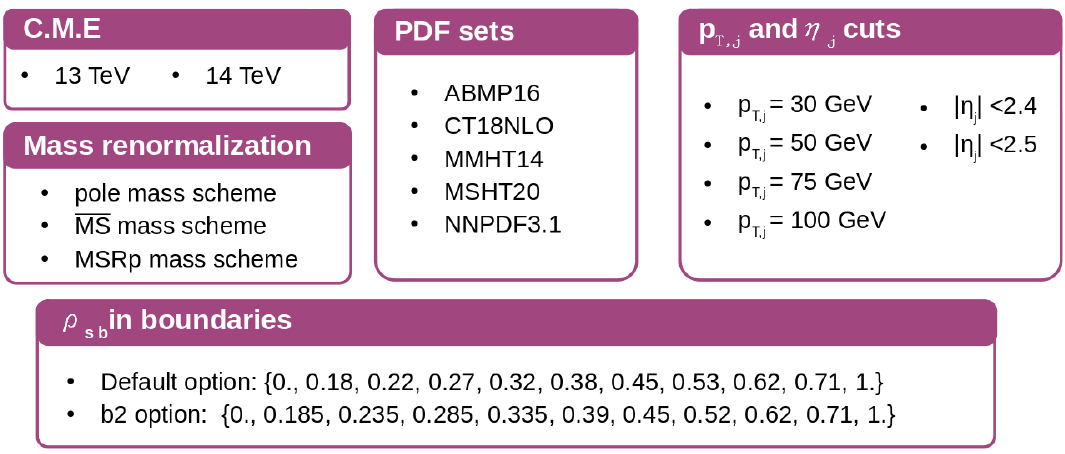}
\caption{\label{fig:input} Input options for the numerical tables of differential cross-sections for $t\bar{t}j + X$ hadroproduction, collected in our website.}
\end{center}
\end{figure}

All the available predictions we produced are collected on a website,\\
\vspace{0.1cm}\\
\centerline{
\texttt{https://ttj-phenomenology.web.cern.ch/}
}
\vspace{0.1cm}\\
from where they can be downloaded as tables of numerical values,
ready for use in the experimental analyses or for further
phenomenological studies. The tables correspond to different analysis
cuts and center-of-mass energies ($\sqrt{S}$ = 13 and 14 TeV), 
and include predictions in different top-quark mass renormalization schemes, 
for different values of $m_t^R$ in steps of 1 GeV. 
Results for different scales $\mu_0$ are reported. The results for scales 
\{(1,1), (2,2), (0.5, 0.5)\} $\mu_0$ are tabulated separately, allowing the user to build a scale uncertainty band. 
Results for different PDF sets are also tabulated separately.
Two $\rho_s$ binning choices are considered, reflecting the most recent bin optimization efforts by the ATLAS and CMS collaboration.  
A schematic of the various input options used for building the tables is presented in Fig.~\ref{fig:input}.
\\
\\

Additionally, the website includes a number of plots,
showing the behaviour of the total fiducial cross sections 
as a function of the $\mu_R$ and $\mu_F$ scales
after the cuts specified in subsection~\ref{subsec:cuts}, the $\rho_s$, $\mathcal{R}$ and also many other differential distributions for various inputs, 
as described in subsection~\ref{subsec:input}, together with their uncertainties. Further tables of numerical values and additional plots can be produced on request. 

\subsection{Selected results}

In the following we report selected results illustrative of the behaviour of the
$\rho_s$ and $\mathcal{R}$ distributions as a function of various
inputs, as described in subsection~\ref{subsec:input}, under the cuts listed in subsection~\ref{subsec:cuts}. 

\subsubsection{$\rho_{\mathrm{s}}$ and $\mathcal{R}$ distributions: scale variation uncertainty}
\begin{figure}[htb]
\centerline{%
\includegraphics[width=12.5cm]{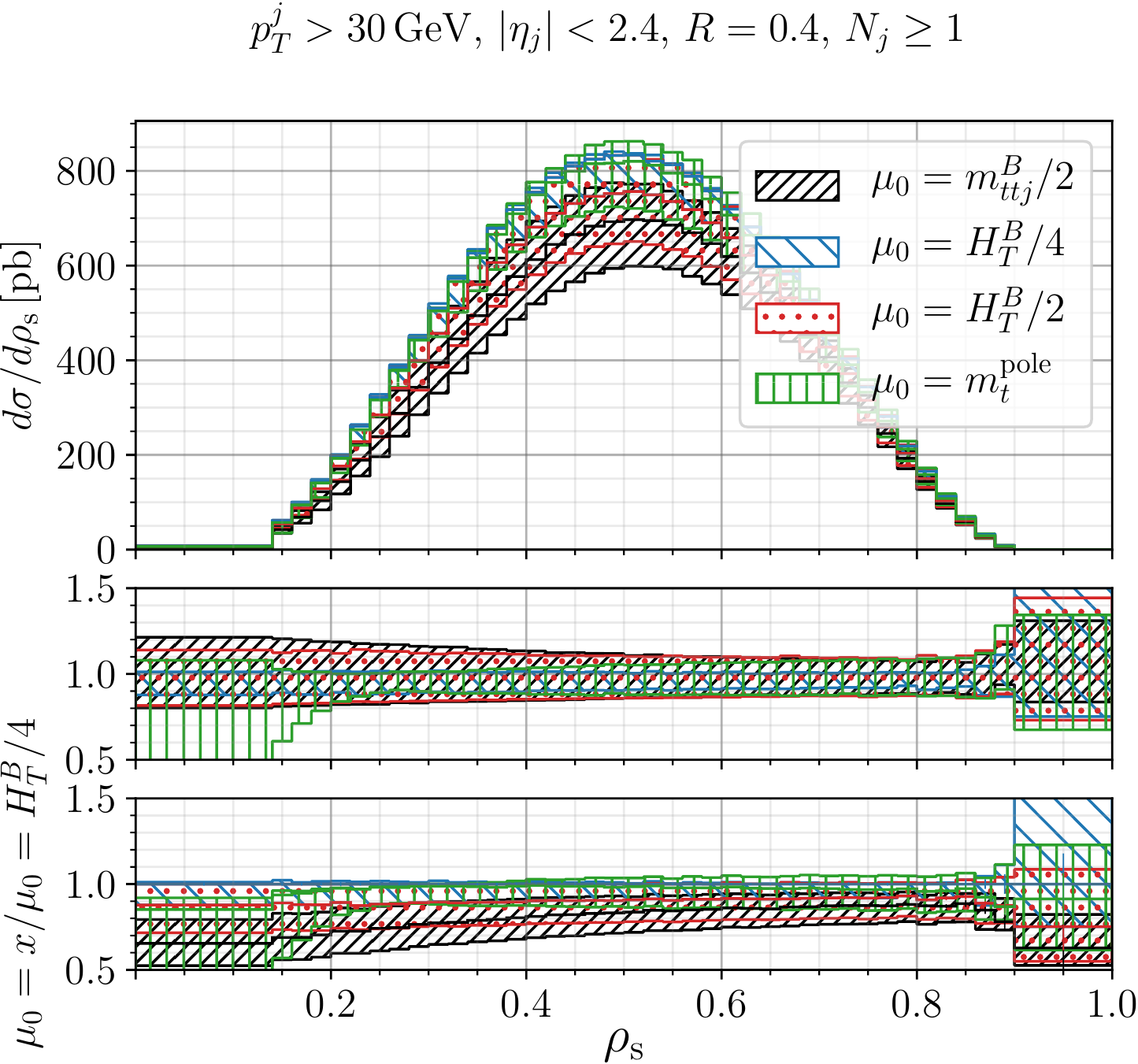}}
\caption{NLO differential cross section of the process $pp \rightarrow t\bar{t}j +X$ at $\sqrt{s}=13\,$TeV as a function of $\rho_\mathrm{s}$ obtained with the scales $\mu_0=m_t^{\mathrm{pole}}$ (green), $m_{t\bar{t}j}^B/2$ (black), $H_T^B/2$ (red) and $H_T^B/4$ (blue). The scale variation uncertainty bands are obtained by taking the envelope of the seven-point scale variation graphs, while the prediction obtained with $K_R = K_F = 1$ is shown as solid line.}
\label{Fig:rho_scale}
\end{figure}

As the $\rho_{\mathrm{s}}$ distribution is particularly interesting in view of the top-quark mass extraction, we investigate the influence of the scale definition on the scale variation uncertainty. Thereby an inappropriate central scale definition can lead to large higher-order corrections and large scale variation uncertainties. NLO QCD predictions for the four scale choices $\mu_0=m_t$, $m_{t\bar{t}j}^B/2$, $H_T^B/2$ and $H_T^B/4$, together with the corresponding seven-point scale-variation uncertainties, are shown in Fig.~\ref{Fig:rho_scale}.
In the ratio plot of the middle panel all scale variation bands are rescaled by the corresponding central scale prediction, to allow for a direct comparison of the size of the scale uncertainty obtained with different central scales. To further visualize the agreement within scale uncertainty between the NLO predictions obtained with these different scale choices, in the lower ratio plot all scale variation uncertainty bands are rescaled by the central scale prediction obtained with $\mu_0=H_T^B/4$. As shown in Fig.~\ref{Fig:rho_scale}, the scale variation uncertainty increases rapidly for low values of $\rho_s$, corresponding to large values of $m_{t\bar{t}j}$, when applying the static scale $\mu_0=m_t$. This signals that the corresponding phase-space region is not well described using the static scale choice. For $0.1 \lesssim \rho_\mathrm{s} \lesssim 0.3$ the scale uncertainty using $\mu_0=m_t$ is strongly reduced and of comparable size to the one obtained with the dynamical scale $\mu_0=H_T^B/4$, which shows the smallest scale uncertainty over the whole phase-space region, excluding the last bin, in which only low statistics is available. This strong reduction of the scale uncertainty in this phase-space region using $\mu_0=m_t$ was found to be artificial and due to the crossing of the scale variation graphs. In general, the description using either central scale is more similar in the bulk of the $\rho_{\mathrm{s}}$ distribution compared to the high energy tail. 

\begin{figure}[htb]
\centerline{%
\includegraphics[width=12.5cm]{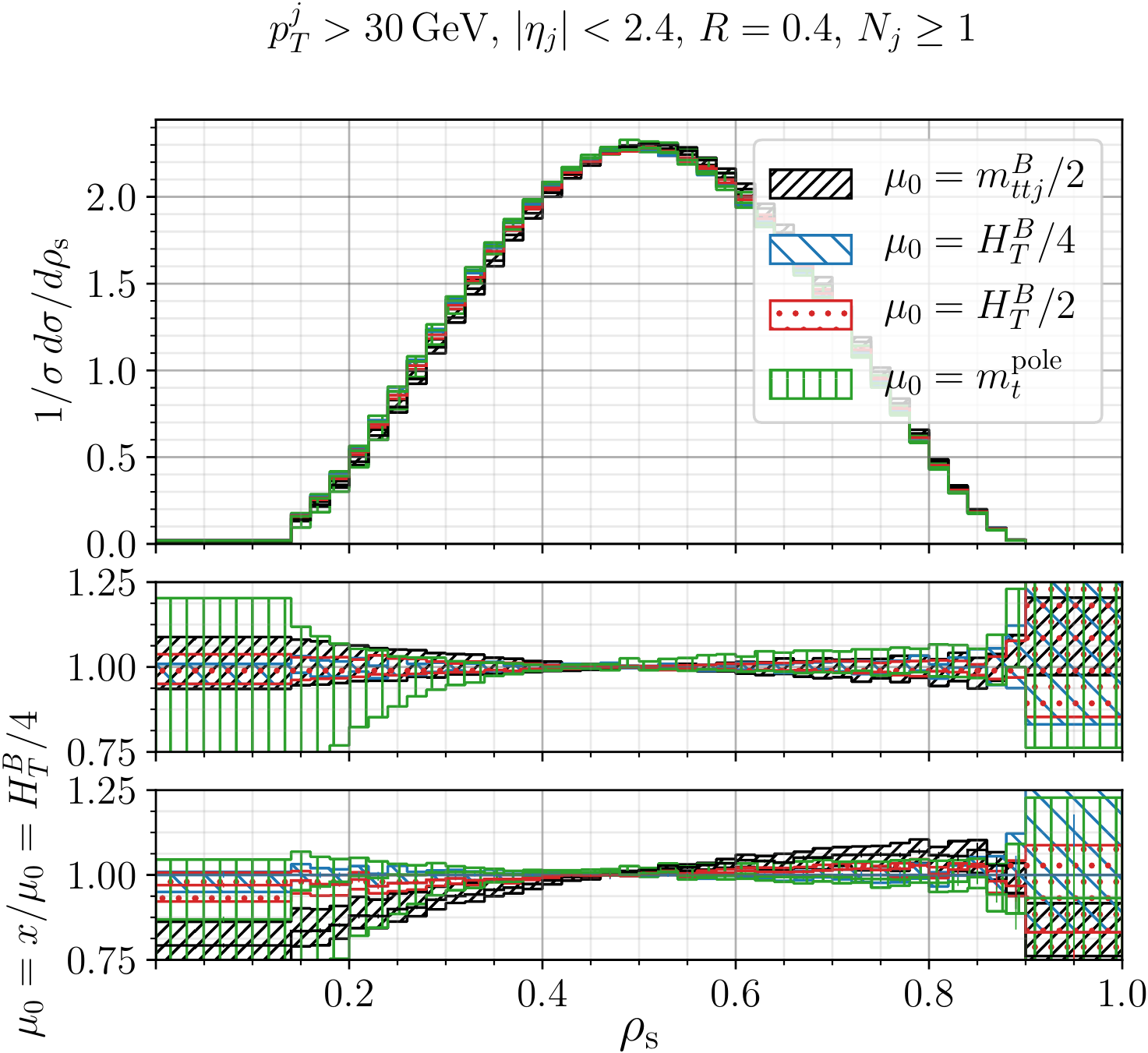}}
\caption{Same as Fig.~\ref{Fig:rho_scale}, but for the normalised NLO $\rho_{\mathrm{s}}$ distribution (i.e. $\mathcal{R}$ distribution).}
\label{Fig:rho_norm_scale}
\end{figure}

As for the top-quark mass extraction the normalized $\rho_{\mathrm{s}}$ distribution is of particular interest, it is shown in Fig.~\ref{Fig:rho_norm_scale}. 
The strong reduction of the scale uncertainty when using a dynamical instead of the static scale $\mu_0=m_t$ was found to originate from the shape variation of the $\rho_\mathrm{s}$ distribution induced by scale variation. When using a dynamical scale, the seven-point scale variation graphs exhibit a similar shape, but different normalizations. On the other hand, in case of the static scale, also significant shape distortions of the $\rho_{s}$ distribution are induced by scale variation. This leads to significantly enhanced scale variation uncertainties in the low $\rho_\mathrm{s}$ tails and slightly enhanced scale variation uncertainties in the bulk of the normalised $\rho_\mathrm{s}$ distribution when using the static instead of the considered dynamical scales. 

\subsubsection{$\mathcal{R}$ distribution: PDF uncertainties}
\begin{figure}[htb]
\centerline{%
\includegraphics[width=12.5cm]{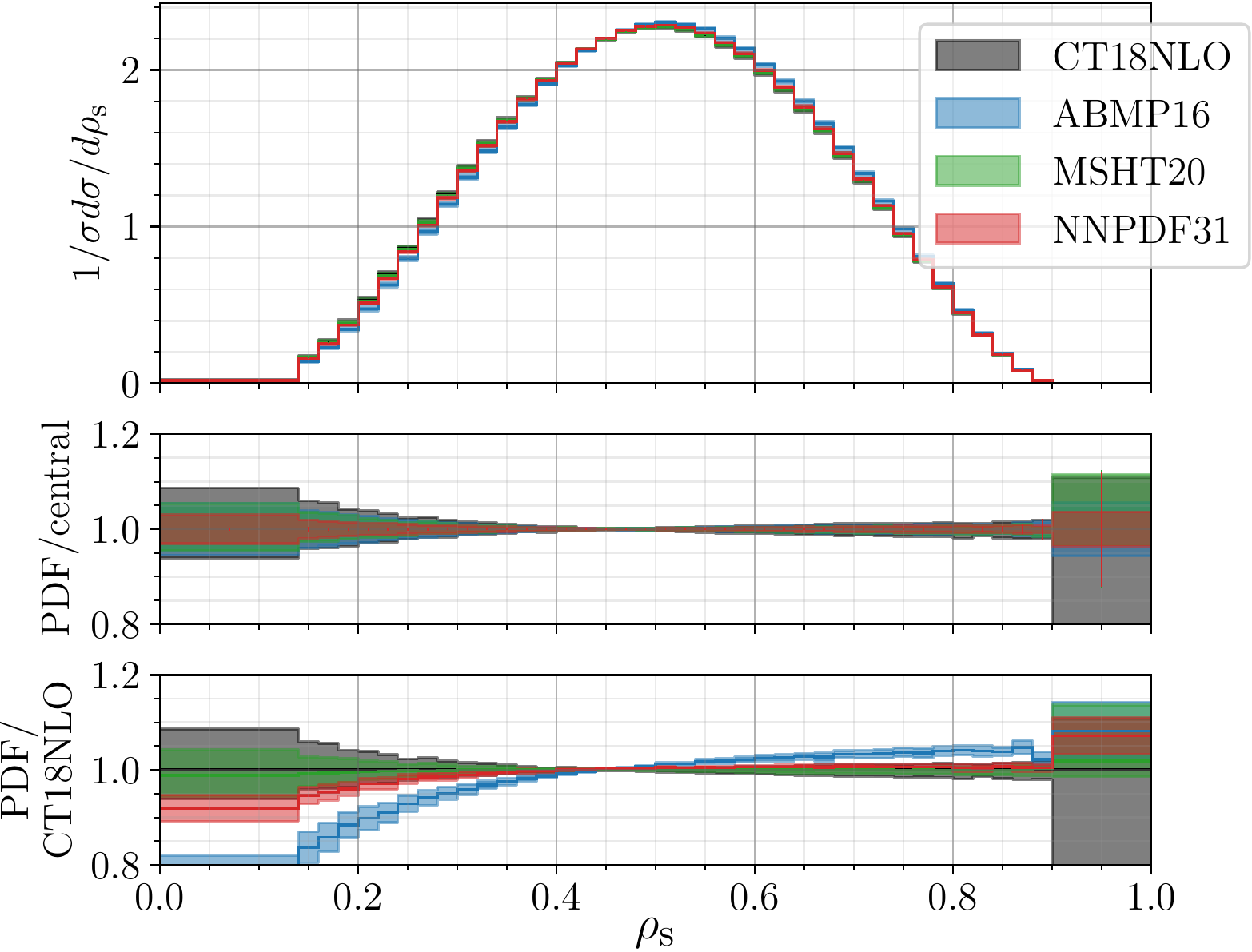}}
\caption{Approximate NLO PDF uncertainties for the 
normalized $\rho_\mathrm{s}$, i.e. $\mathcal{R}$, distribution of the process $pp \rightarrow t\bar{t}j +X$ at $\sqrt{s}=13$~TeV obtained in a computation with LO matrix-elements and NLO PDFs and $\alpha_S$ evolution, using as input the dynamical scale $\mu_0=H_T^B/4$ and the CT18NLO (black), ABMP16 (blue), MSHT20 (green) and NNPDF3.1 (red) NLO PDF sets. Each PDF uncertainty is calculated as recommended by the authors of the corresponding PDF fit. The CT18NLO PDF uncertainty rescaled from the 90\% Confidence Level to the 68\% Confidence Level as provided by the other PDF fits.}
\label{Fig:rho_norm_pdf}
\end{figure}

In Fig.~\ref{Fig:rho_norm_pdf} the PDF uncertainty in the normalised $\rho_{\mathrm{s}}$ distribution, i.e. $\mathcal{R}$ distribution, calculated by applying the scale definition $\mu_0=H_T^B/4$, is shown for four different PDF sets, including our default PDF set CT18NLO~\cite{Hou:2019efy} (black) and additionally the predictions obtained with ABMP16~\cite{Alekhin:2018pai} (blue), MSHT20~\cite{Bailey:2020ooq} (green) and NNPDF3.1~\cite{NNPDF:2017mvq} (red) NLO PDF sets. Thereby an approximation was applied, in which the NLO PDF uncertainty is calculated by using the LO hard scattering amplitudes for computing partonic cross-sections and a NLO PDF set. This reduces the required computational effort in terms of CPU hours, and facilitates the computation of PDF uncertainties, considering that the latter require to make runs with many different sets for each PDF choice. This procedure was validated by comparing its results with those of full NLO calculations. It turned out that this approach can be applied because it leads to PDF uncertainties whose relative size is very similar to that obtained in a full NLO calculation. In the ratio plot in the middle panel each PDF uncertainty band is rescaled by the prediction obtained with the respective central PDF set. To visualize again the relative differences of the predictions, in the ratio plot in the lower panel all PDF uncertainty bands are rescaled by the prediction obtained with the central CT18NLO PDF set. 
In the low $\rho_s$ tail the predictions calculated with different PDF sets show the largest discrepancy among each other, which are not covered by the PDF uncertainty bands. This was found to be due to the different behaviour of the gluon PDFs at large Bjorken-$x$. For events leading to an $\rho_{\mathrm{s}}$ value in the first bin, i.e. $\rho_\mathrm{s} \in [0,0.14]$, corresponding to high $t\bar{t}j$ invariant-mass tail, the minimal momentum fraction $x_{\mathrm{min}}$ and maximal momentum fraction $x_{\mathrm{max}}$ distributions are peaked around $x_{\mathrm{min}}=0.15$ and $x_{\mathrm{max}}=0.25$, respectively. These distributions are defined through the momentum fractions ($x_1$,$ x_2$) of the incoming partons as $x_{min} = \mathrm{min}(x_1,x_2)$ and $x_{max} = \mathrm{max}(x_1,x_2)$. Considering instead the phase-space region in the bulk of the $\rho_{\mathrm{s}}$ distribution, peaks of the distributions are found at $x_{min}=0.02$ and $x_{max}=0.07$. In this region of lower $x$-values the gluon PDFs of the different PDF fits are in much better agreement than for ($x_1$, $x_2$) $> 0.1$. 
New QCD analyses focused on the extraction of improved PDFs at large $x$ using forthcoming high-statistics LHC experimental data, as well as future data at the Electron-Ion Collider~\cite{AbdulKhalek:2021gbh}, will be crucial to pin down the large-$x$ PDF uncertainties. 
  \\

\subsubsection{$\rho_s$ and $\mathcal{R}$ distributions: perturbative convergence of the predictions and comparison of uncertainties from different input sources}

\begin{figure}[htb]
\centerline{%
\includegraphics[width=0.48\textwidth]{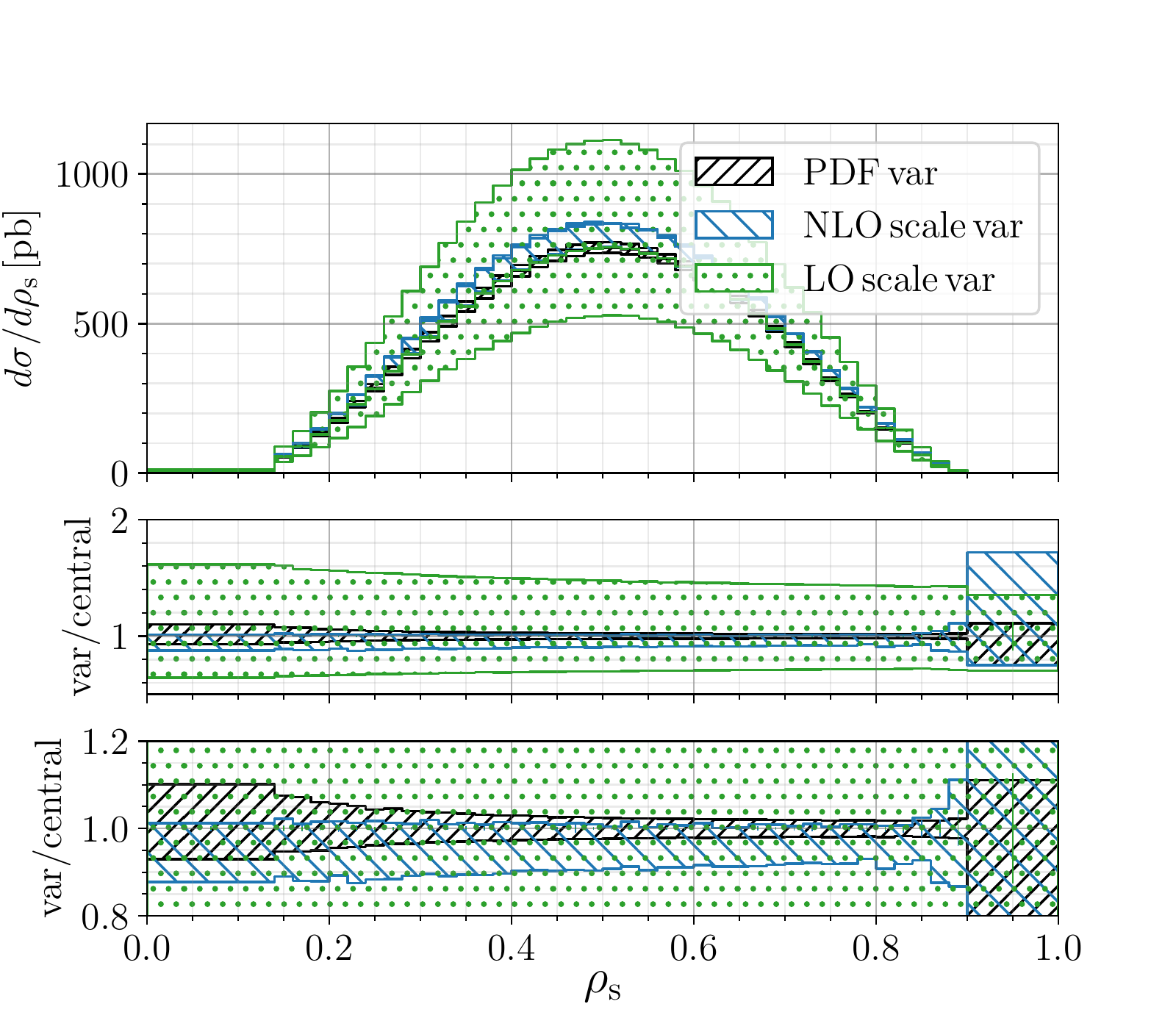}
\includegraphics[width=0.48\textwidth]{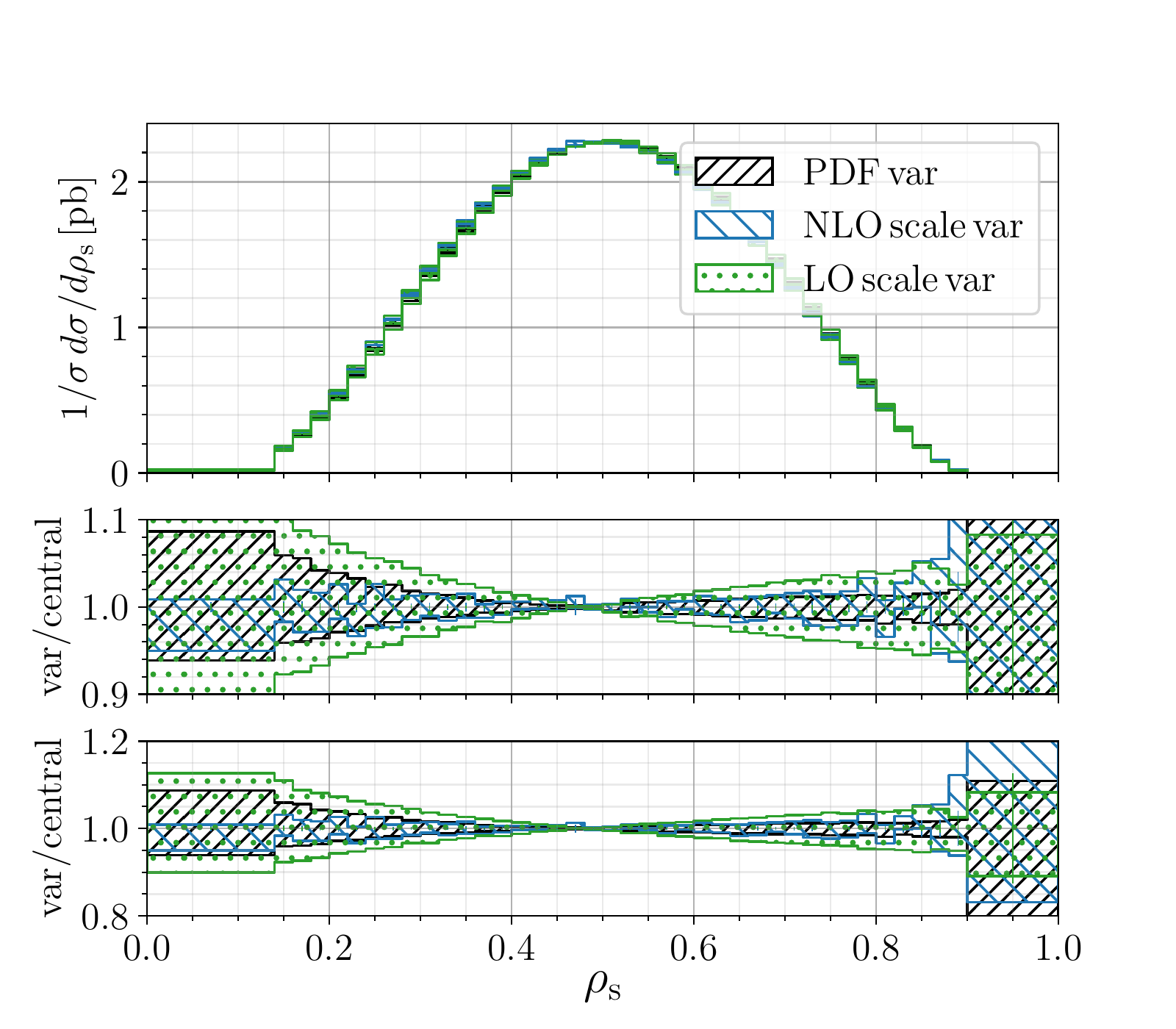}}
\caption{NLO (blue) and LO (green) scale variation uncertainty compared to the approximate PDF uncertainty  bands (black) obtained using the dynamical scale $\mu_0=H_T^B/4$ and CT18NLO are shown for the $\rho_{\mathrm{s}}$ (left) and the normalised $\rho_{\mathrm{s}}$ distribution (right) of the process $pp \rightarrow t\bar{t}j$ at $\sqrt{s}=13\,$TeV.}
\label{Fig:rho_pdf_scale}
\end{figure}

Additionally, we show in Fig.~\ref{Fig:rho_pdf_scale} the comparison of the scale and PDF uncertainties obtained using as input the CT18NLO PDF set and the dynamical scale $\mu_0=H_T^B/4$ in the $\rho_{\mathrm{s}}$ distribution (left) and the normalized $\rho_{\mathrm{s}}$, i.e. $\mathcal{R}$, distribution (right). When applying the dynamical scale choice $\mu_0=H_T^B/4$, the NLO scale variation uncertainty turns out to be strongly reduced when contrasted with the static scale choice and even comparable to the PDF uncertainty in the normalized distribution. In Fig.~\ref{Fig:rho_pdf_scale} also the LO scale uncertainty band is shown and the improved perturbative stability reached with the NLO prediction is clearly visible in the relative size of the NLO and LO scale uncertainty bands.\\
Additional studies regarding the scale choice included the comparison of the NLO and LO scale variation bands for a number of differential distributions. A more consistent behaviour was found using the dynamical scales $\mu_0=H_T^B/2$ and $\mu_0=H_T^B/4$. Additionally, it was observed that the value of the $R$ parameter in the anti-$k_T$ jet clustering algorithm, varied in the interval $R$~=~0.4 - 0.8, does not have a large impact on the scale uncertainty, but a larger $R$ parameter leads to larger differential cross sections and thus helps increasing the statistical accuracy of the measurements and top-quark mass extraction. 

\subsubsection{Linearity of $\mathcal{R}$ with the top-quark mass in different renormalization schemes}
\begin{figure}[h!]
\begin{center}
\includegraphics[width=0.48\textwidth]{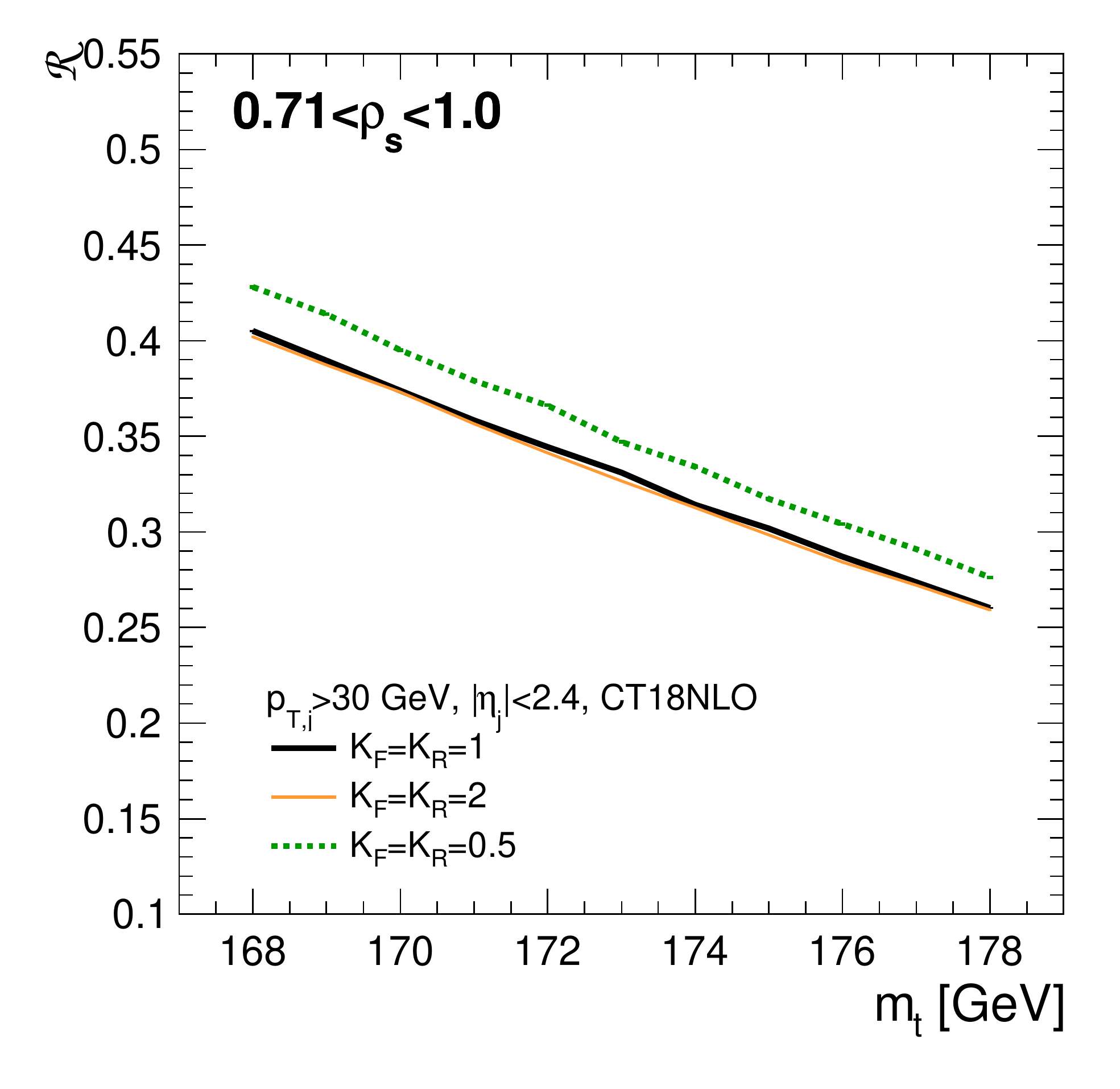}
\includegraphics[width=0.48\textwidth]{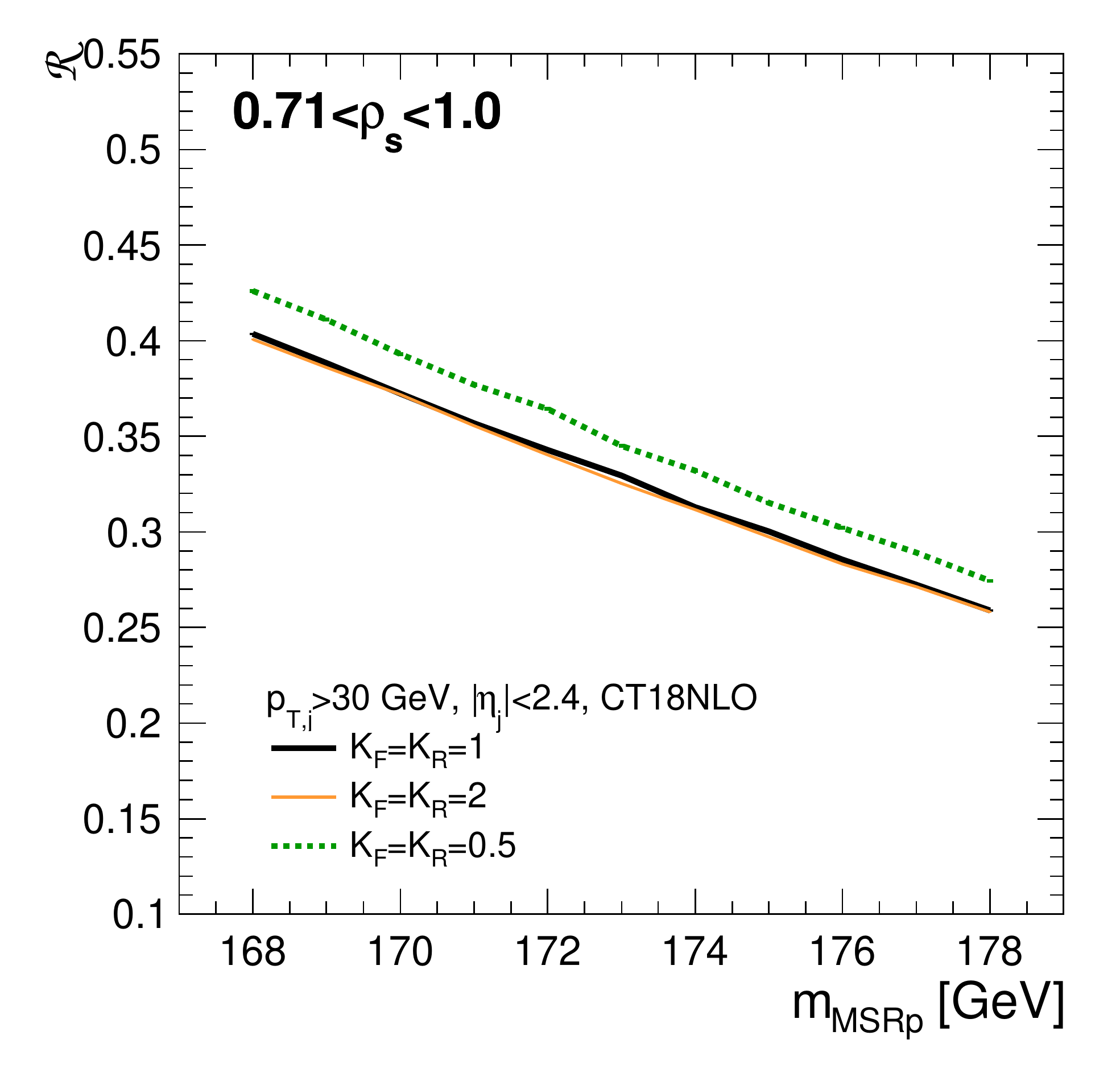}
\includegraphics[width=0.48\textwidth]{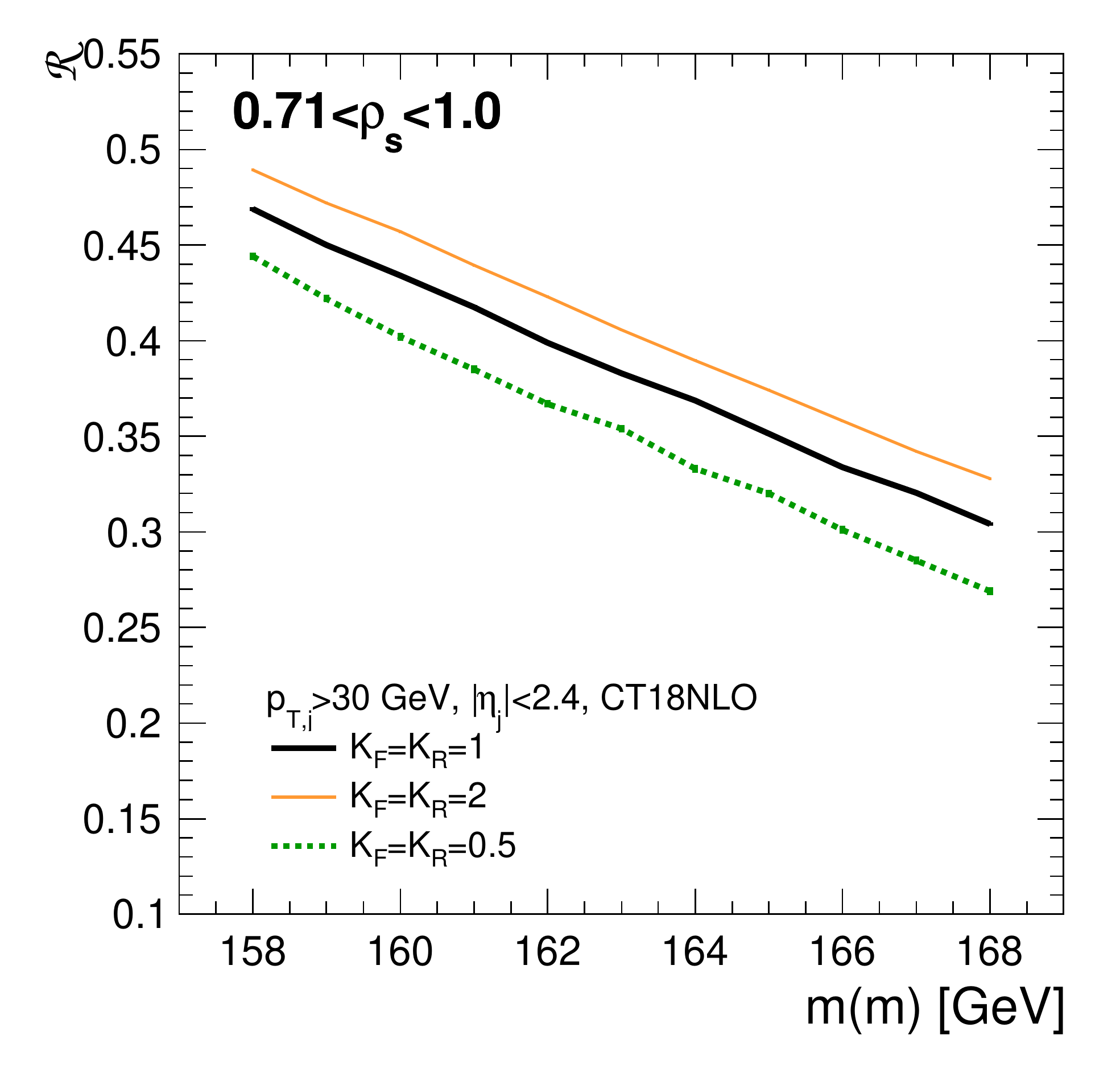}
\caption{
\label{fig:linear}
$\mathcal{R}$ distribution in the large $\rho_s$ bin [0.71,1]
as a function of $m_t^R$, evaluated with top-quark mass renormalized 
in the pole (upper left panel), MSR (upper right panel) and
 and ${{\rm \overline{MS}}}$ (lower panel)
scheme, respectively. Central predictions evaluated with
$\mu^0 = m_t^R$ are plotted together with those from  scale variation
(2, 2)$\mu_0$ and (0.5, 0.5)$\mu_0$. 
 } 
\end{center}
\end{figure}

The behaviour of the $\mathcal{R}$ distribution in the large $\rho_s$ bin [0.7, 1] as a function of the top-quark mass value is plotted in Fig.~\ref{fig:linear}.
Different panels refer to computations in different top-quark mass renormalization schemes (on-shell, MSR and ${{\rm \overline{MS}}}$). 
In all schemes, it is evident that $\mathcal{R}$ decreases linearly with increasing top-quark mass value, for top-quark mass values in a wide interval, ranging from 168 to 178 GeV. This behaviour induced us to report tabulated results for the $\mathcal{R}$ distribution in each mass renormalization scheme for ten fixed values of the top-quark mass, in 1~GeV steps. The $\mathcal{R}$ distribution for any intermediate top-quark mass value in the aforementioned interval can be simply obtained by linear interpolation of the predictions in adjacent bins. Comparing the results of scale variation in the three different schemes, it is evident that the behaviour of predictions in the on-shell and MSR schemes is quite similar, whereas predictions in the ${{\rm \overline{MS}}}$ scheme are accompanied by larger scale uncertainty. This is related to the shortcomings of the ${{\rm \overline{MS}}}$ scheme in the threshold region, corresponding to the largest
$\rho_s$ values. Considering the shortcomings of the on-shell scheme related to renormalon ambiguity already discussed in subsection~\ref{subsec:input}, we can conclude that the MSR scheme is a viable alternative, and we encourage future top-quark mass extractions in this scheme. Our predictions~\cite{Alioli:2022lqo} are the first ones for $t\bar{t}j+X$ ever published in this scheme, after first esults on MSR cross-sections for $t\bar{t}$ hadroproduction that we presented in Ref.~\cite{Garzelli:2020fmd}.

\section{Summary, observations and recommendations for future analyses}
\label{sec:conclu}

We have performed NLO QCD studies of the $\rho_s$ and $\mathcal{R}$ distributions, useful to extract the top-quark mass values from analyses of the ongoing and forthcoming $pp \rightarrow t\bar{t}j + X$ LHC experimental data.  
As discussed in Section~\ref{sec:pheno}, we have computed a comprehensive set of predictions using a wide set of inputs.
The results are publicly available from a website, organizing them in hundreds of numerical tables and plots.
In this document, we have presented selected results and discussed the most relevant uncertainties affecting them. Further results and discussions on the topic can be found in Ref.~\cite{Alioli:2022lqo}. 

Present experimental analyses are especially focused on relatively large $\rho_s$ values. The progressive accumulation of high-statistics experimental data will make possible to extend the $\rho_s$ interval, progressively covering more extreme (i.e. lower and larger) $\rho_s$ values, hence exploiting the mass sensitivity on a broader range of $\rho_s$. However, using the forth\-coming high-statistics data in an extended $\rho_s$ interval requires high-accuracy predictions, deep understanding of the perturbative behaviour of the calculation and of its dependence on further inputs like $\alpha_s(M_Z)$, the PDFs and the jet reconstruction procedure. Systematic studies have been performed in Ref.~\cite{Alioli:2022lqo} and recapitulated in this document.
In the following, we summarize our main observations and recommendations: 

\begin{itemize}

\item 
in the computation of the $\mathcal{R}$ distribution use dynamical scales: in particular the choice $\mu_0 = H_T/4$ turns out to be particularly interesting because of the perturbative convergence and minimization of the size of scale uncertainties. While the static scale $\mu_0 = m_t^R$ can still be regarded as a good choice for large $\rho_s$ values, the aforementioned dynamical scale choice is proven to perform definitely better in case of $\rho_s < 0.4$.

\item use state-of-the-art PDF fits, with particular attention to the large-$x$ region. This region, where present PDFs are still quite uncertain, is in fact definitely spanned when computing predictions at small $\rho_s$ values. 

\item use short-distance top-quark mass renormalization schemes, free of renormalon ambiguities, as a viable alternative to the on-shell scheme. At large $\rho_s$, where threshold effects become relevant, the ${{\rm \overline{MS}}}$ scheme does not prove to be competitive, whereas the MSR scheme is expected to be a viable choice over the whole $\rho_s$ range, both in line of principle and according to our first results. Our predictions for $t\bar{t}j + X$  cross-sections in the MSR scheme represent the first example of calculations in this scheme for this process. More studies and analyses on the systematics and potential benefits (and/or shortcomings) inherent the extraction of the top-quark mass in this scheme from $t\bar{t}j + X$ events are indeed welcome.


\item develop methodologies to go beyond NLO accuracy in predictions 
for $t\bar{t}j + X$ (differential) cross-sections with stable top quarks: including NNLO radiative corrections and the effects of resummation of different kinds of large logarithms might be important especially in those regions where scale uncertainties are particularly large. 

\item further refine the experimental top-quark reconstruction procedures. We expect that matching calculations for $ttj + X$ production with full off-shell effects to parton shower approaches will be particularly useful in this respect. 

\end{itemize}

In order to facilitate analyses following these directions, we plan to go on keeping up-to-date our website of $t\bar{t}j + X$ predictions, in such a way to reflect the latest theoretical and experimental developments from both our and other groups.

\section*{Acknowledgments}
We are grateful to the Snowmass community and to many colleagues who
participated in the activities of the EF03 group, for useful discussions
on the top-quark mass topic and methodologies to extract it.  

The work of S.A. and A.G. is supported by the ERC Starting Grant REINVENT-714788. S.A. also acknowledges funding from MIUR through the FARE grant R18ZRBEAFC and from Fondazione Cariplo and Regione Lombardia, grant 2017-2070.
J.F. and A.I. ack\-now\-led\-ge support from projects PGC2018-094856-B-100 (MICINN/FEDER), PROMETEO-2018/060 and CIDEGENT/2020/21 (Generalitat Valenciana), and the iLINK grant LINKB\-20065 (CSIC). The work of M.V.G., S.M. and P.U. was supported in part by the Bundesministerium f\"ur Bildung and Forschung under contracts 05H21GUCCA and 05H18KHCA1.


\bibliographystyle{JHEP}
\bibliography{Cetology}  
\end{document}